\documentclass[paper=A4,11pt,DIV=12,headings=big]{scrartcl}
\pdfoutput=1
\usepackage{graphicx}
\usepackage{tabularx}
\usepackage{array}
\usepackage{amsmath}
\usepackage{amssymb}
\usepackage{xcolor}
\usepackage{longtable}
\usepackage{verbatim}
\usepackage{cite}
\usepackage{bbm}
\usepackage{multirow}
\usepackage{enumerate}
\usepackage[normalem]{ulem}
\definecolor{darkblue}{rgb}{0,0,0.5}
\usepackage[colorlinks,linkcolor=darkblue,citecolor=darkblue,urlcolor=darkblue,%
]{hyperref}

\allowdisplaybreaks

\addtokomafont{disposition}{\rmfamily\boldmath}
\newcommand{\mailref}[1]{\href{mailto:#1}{#1}}

\begin{document}

\noindent FLAVOUR(267104)-ERC-67
\vspace{1cm}

\begin{center}
{\Large
\bf\boldmath
Flavour physics  and flavour symmetries\\[0.2cm] after the first LHC phase}
\\[0.8 cm]
{\large
R.~Barbieri$^{a}$, D.~Buttazzo$^{b}$, F.~Sala$^{c}$, and D.\,M.~Straub$^{d}$} \\[0.4 cm]
\small
$^a$ {\em Scuola Normale Superiore and INFN, Piazza dei Cavalieri 7, 56126 Pisa, Italy}\\[0.1cm]
$^b${\em TUM Institute for Advanced Study, Lichtenbergstr. 2a, 85747 Garching, Germany} \\[0.1cm]
$^c${\em Institut de Physique Th\'eorique, CNRS and CEA/Saclay, 91191 Gif-sur-Yvette, France} \\[0.1cm]
$^d${\em Excellence Cluster Universe, TUM, Boltzmannstr.~2, 85748~Garching, Germany} \\[0.4cm]
E-mail: \mailref{barbieri@sns.it}, \mailref{dario.buttazzo@tum.de}, \mailref{filippo.sala@cea.fr}, \mailref{david.straub@tum.de}%
\end{center}

\smallskip
%###################################################################
\begin{abstract}\noindent
Based on flavour symmetries only, there are two ways to give rise to an effective description  of flavour physics in the quark sector close to the CKM picture: one is based on  $U(3)_q\times U(3)_u\times U(3)_d$ (or equivalent)  and the other on   $U(2)_q\times U(2)_u\times U(2)_d$ (or equivalent). In this context we analyze the current status of flavour physics measurements
 and we compare their impact, in the specific case of supersymmetry, with the direct searches of new particles at the LHC, present or foreseen.
\end{abstract}
%###################################################################

%###################################################################
\section{Introduction} \label{sec:Intro}
%###################################################################
Which  impact can flavour measurements  have as a whole on the medium term future of particle physics? There are at least two different ways to try to address this question. One way is to consider the highest energies that flavour measurements can probe by taking a blind attitude on the possible structure of flavour physics: as well known these energies can extend in some cases up to $10^{4\div 5}$ TeV. Opposite to this is the view that new flavour physics phenomena may be related to Beyond the Standard Model (BSM) physics in the TeV range, if there is any, still compatible with current  LHC bounds. In this work we  reexamine the latter case.

 The current  overall status of flavour measurements in the quark sector shows consistency with the Cabibbo-Kobayashi-Maskawa (CKM) picture of flavour physics with an uncertainty  at 20 to 30\% level (see below). From an effective field theory point of view the CKM picture, as embedded in the Standard Model (SM), amounts  to the presence of a limited number of operators with a flavour dependence accounted for by a specific combination of CKM matrix elements. In view of this, it is interesting to consider an effective picture close enough to the CKM one, as just defined, based on symmetries and on their possible breaking pattern only, 
 without referring to specific models or to appropriately chosen dynamical assumptions. The well known example goes under the name of Minimal Flavour Violation (MFV) \cite{Chivukula:1987py,Hall:1990ac,DAmbrosio:2002ex} and is based on the flavour symmetry in the quark sector of the SM
  \begin{equation}
 U(3)_q\times U(3)_u \times U(3)_d\equiv U(3)^3,
 \end{equation}
 only broken by the Yukawa matrices transforming under  this group, like spurions, as 
 \begin{equation}
 Y_u = (3, \bar{3},1),\quad\quad Y_d = (3, 1, \bar{3}).
 \label{YuYd}
 \end{equation}
 
  As we are going to illustrate, at least within continuous symmetries, there seems to be effectively a single other CKM-like pattern, as arising from\cite{Barbieri:2012uh,Barbieri:2011ci}
  \begin{equation}
  U(2)_q\times U(2)_u\times U(2)_d\equiv U(2)^3,
  \end{equation}
   broken by
  \begin{equation}
 \Delta_u = (2, \bar{2},1),\quad\quad \Delta_d = (2, 1, \bar{2}),\quad\quad \mathcal{V} = (2, 1, 1).
 \label{U2^3}
 \end{equation}
$U(2)^3$ is the flavour symmetry of the SM in the quark sector if only the top and the bottom Yukawa couplings $y_t$ and $y_b$ are kept. For this reason, unlike $U(3)^3$, $U(2)^3$ is an approximate symmetry of the observed quark flavour structure with all the parameters in eq.  (\ref{U2^3}) small, at a few $10^{-2}$ or less. While  $U(2)^3$ has been considered in the literature as  a limiting cases of $U(3)^3$  broken by the parameters in eq. (\ref{YuYd}) when both $y_t$ and $ y_b$ are large \cite{Feldmann:2008ja,Kagan:2009bn}, hence large $\tan{\beta}$, this is not a logical necessity.

In a specific model of new physics at the TeV scale, there may be different ways to make it compatible with flavour physics measurements. Nevertheless, the generality of the approach described so far makes relevant to compare it with the current status of flavour physics measurements, as we do in the first part of the paper.
In the second part of the paper we shall make a detailed comparison of flavour measurements with the direct searches of new particles in the case of supersymmetry, supplemented with the dynamical assumption that the stops and the sbottoms are generally lighter than squarks of  the first and second generations. Within this framework, we also quantify the strong impact of the recent upper bound on the electron EDM.

\section{Effective operators}
\label{eff-ops}

For ease of the reader we recall the general structure of the effective operators that result in the two different CKM-like patterns.

Let us consider first the $U(3)^3$ case, which leads to  MFV. By uninfluential transformations inside the flavour symmetry group
one can diagonalize the up quark mass matrix, so that the only non-trivial flavour structure comes from the diagonalization of the down quark mass matrix. It then follows, working to first order in all the flavour breaking parameters except $y_t$, that the only quark bilinears that are consistent with the symmetries and may produce a breaking of  flavour after going to the physical quark bases are
\begin{equation}
\bar{\mathcal{Q}}I_3\gamma_\mu \mathcal{Q},\quad\quad \bar{\mathcal{Q}}I_3 Y_d(1, \sigma_{\mu \nu})\mathcal{D}
\label{bili}
\end{equation}
where $\mathcal{Q}$ and $\mathcal{D}$ are triplet vectors in family space,  $I_3$ is the diagonal matrix with  only the $33$ element non zero and $Y_d$ is the down Yukawa matrix. In  turn, by going to the physical bases, this  leads to the following set of relevant operators $(\xi_{ij}=V_{tj}V_{ti}^*$ where $V$ is the CKM matrix):
\begin{enumerate}[i)]
\item $\Delta F =2$:
\begin{equation}
C_{LL}  \xi_{ij}^2 \frac{1}{2}(\bar{d}_{Li}\gamma_\mu d_{Lj})^2,
\label{CLL}
\end{equation}
\item $\Delta F =1$, chirality breaking ($\alpha = \gamma, G$):
\begin{equation}
C^\alpha e^{i \chi^\alpha} \xi_{ij}  (\bar{d}_{Li}\sigma_{\mu \nu} m_j d_{Rj})  O^\alpha_{\mu \nu}, ~~
O^\alpha_{\mu \nu}= e F_{\mu\nu}, ~ g_s G_{\mu \nu},
\end{equation}
\item $\Delta F =1$, chirality conserving ($\beta = L, R, H$):
\begin{equation}
C^\beta  \xi_{ij}(\bar{d}_{Li}\gamma_\mu d_{Lj})O^\beta_\mu, ~~
O^\beta_\mu = (\bar{l}_L\gamma_\mu l_L),~(\bar{e}_R\gamma_\mu e_R),~(H^\dagger D_\mu H).
\end{equation}
\end{enumerate}
In an effective Lagrangian approach, every operator is meant to be weighted  by the square of an inverse mass scale $1/\Lambda$. The numerical coefficients $C_{LL}, C^\alpha, C^\beta$ are of order unity and real and the only phases $\chi^\alpha$ are made explicit in the chirality breaking operators.

The same effective operators are present in $U(2)^3$, except that more phases and  relative factors of order unity  between the third and the first two generations are introduced\cite{Barbieri:2012uh}. This comes about because of the reduced flavour group and the presence in eq. (\ref{U2^3}) of the doublet $ \mathcal{V}$. It is no longer possible by transformations inside the flavour group to diagonalize the up quark mass matrix. Nevertheless, both the up and the down quark mass matrices, always working to first order in all the breaking parameters except $y_t$, can be diagonalized by pure  transformations of the left handed fields only.
Furthermore the presence of $ \mathcal{V}$, introduces new quark bilinears other than those in eq. (\ref{bili}),  that lead to a distinction between the third and the first two families other than the one contained in the CKM matrix. An example is $\bar{Q}\mathcal{V} \gamma_\mu q_3$ with $Q$ a doublet under $U(2)_q$. Consequently the explicit structure of the relevant effective operators in $U(2)^3$ is:
\begin{enumerate}[i)]
\item $\Delta B =2, i = s,d$:
\begin{equation}
c_{LL}^B e^{i \phi_B} \xi_{ib}^2 \frac{1}{2}(\bar{d}_{Li}\gamma_\mu b_L)^2,
\label{eff1}
\end{equation}
\item $\Delta S =2$:
\begin{equation}
c_{LL}^K  \xi_{ds}^2 \frac{1}{2}(\bar{d}_{L}\gamma_\mu s_L)^2,
\label{eff2}
\end{equation}
\item $\Delta B =1, i = s,d$, chirality breaking ($\alpha = \gamma, G$):
\begin{equation}
c^\alpha e^{i \phi^\alpha} \xi_{ib} m_b (\bar{d}_{Li}\sigma_{\mu \nu} b_R)  O^\alpha_{\mu \nu}, ~~
O^\alpha_{\mu \nu}= e F_{\mu\nu}, ~ g_s G_{\mu \nu},
\end{equation}
\item $\Delta B =1, i = s,d$, chirality conserving\footnote{There are also four-quark $\Delta F=1$ operators (the analogues of QCD penguins in the SM). While they are not relevant for the observables we consider below, they are relevant for mixing-induced CP asymmetries in non-leptonic penguin decays like $B\to\phi K_S$ or $B\to \eta' K_S$, that can be generated in $U(2)^3$ \cite{Barbieri:2011fc}.} ($\beta = L, R, H$):
\begin{equation}
c^\beta_B e^{i\phi^ \beta} \xi_{ib} (\bar{d}_{Li}\gamma_\mu b_L) O^\beta_\mu, ~~
O^\beta_\mu = (\bar{l}_L\gamma_\mu l_L),~(\bar{e}_R\gamma_\mu e_R),~(H^\dagger D_\mu H),
\end{equation}
\item $\Delta S =1$, chirality conserving:
\begin{equation}
c^\beta_K  \xi_{ds}(\bar{d}_{L}\gamma_\mu s_L)O^\beta_\mu, ~~
O^\beta_\mu = (\bar{l}_L\gamma_\mu l_L),~(\bar{e}_R\gamma_\mu e_R),~(H^\dagger D_\mu H).
\label{eff5}
\end{equation}
\end{enumerate}

\section{Equivalent symmetries or symmetries that do not work}

To the best of our knowledge every other symmetry inside $U(3)^3$ either leads to one of the two patterns described above or deviates from a CKM-like picture because of the occurrence of  different operators and/or of operators dependent on combinations of the CKM matrix elements different from $\xi_{ij}$. 

A case that gives rise to MFV other than  $U(3)^3$ is 
 \begin{equation}
 U(2)_q\times U(2)_u\times U(3)_d\equiv U(2)^2\times U(3),
 \end{equation}
  broken by
  \begin{equation}
 \Delta_u = (2, \bar{2},1),\quad\quad \Delta_d = (2, 1, \bar{3}),\quad\quad\tilde{\Delta}_d = (1, 1, \bar{3}).
 \label{U2_3}
 \end{equation}
That this be true simply follows from the fact that here as well one can diagonalize the up quark mass matrix by transformations inside the symmetry group and that the quark bilinears capable of producing a flavour breaking are the same as in eq. (\ref{bili}). Needless to say, here as in all previous cases, the directions of breaking specified by eq. (\ref{U2_3}) are as important as the symmetry itself. Once they are specified, there is no need to view $U(2)^2\times U(3)$  as a limiting case of $U(3)^3$ broken by the parameters in eq. (\ref{YuYd}), due to the large $y_t$ \cite{Feldmann:2008ja}.

To maintain a CKM-like picture it is essential that one keeps some distinction between left and right fermions. Nevertheless one can for example reduce the $U(2)^3$ group to 
\begin{equation}
U(2)_q\times SU(2)_R\times U(1)_u\times U(1)_d
\end{equation}
broken by
  \begin{equation}
 \tilde{\Delta}_u = (2, 2)_{(-1,0)},\quad\quad  \tilde{\Delta}_d = (2,  2)_{(0,-1)},\quad\quad \tilde{\mathcal{V}} = (2, 1)_{(0,0)}.
 \label{U2^2}
 \end{equation}
Here $SU(2)_R$ acts as a doublet on the first two generations of right handed up-type quarks, of $U$-charge  1, and, independently, on the first two generations of right handed down-type quarks, of $D$-charge  1.
The distinction between $ \tilde{\Delta}_u$ and $ \tilde{\Delta}_d$ introduced by the $U(1)$ factors is needed to prevent non CKM-like chirality breaking operators.
%both in $\Delta C=1$ and in $\Delta S =1$ channels.\footnote{The bounds on non CKM-like chirality-breaking operators in $\Delta C=1$ or $\Delta S=1$ are weaker than the bounds on $\Delta F=2$ chirality-conserving operators. Specifically a model like this without the distinction between $ \tilde{\Delta}_u$ and $ \tilde{\Delta}_d$ and a natural choice for the extra parameters would be consistent with observations for a value of $\Lambda$ non significantly greater than 10 TeV. With this qualification a $U(2)_q\times U(2)_R$ flavour symmetry could be consistent with a Pati-Salam gauge group.}
To first order in these small breaking parameters, it is easy to see that this case leads to the same pattern as $U(2)^3$. Other symmetries that lead to the same pattern are conceivable.

We do not see how to obtain a different quasi-CKM picture of flavour physics by pure symmetry. An example that does not work but is nevertheless interesting is \cite{Barbieri:2010ar}
\begin{equation}
U(1)_{\tilde{B}_1}\times U(1)_{\tilde{B}_2}\times U(1)_{\tilde{B}_3}\times U(3)_d,
\label{U1cube}
\end{equation}
where  $\tilde{B}_i$ acts on $\mathcal{Q}_i, u_i$ as baryon number, broken by
\begin{equation}
\Delta_1 = \bar{3}_{(-1,0,0)},\quad\quad\Delta_2 = \bar{3}_{(0,-1,0)},\quad\quad\Delta_3 = \bar{3}_{(0,0,-1)}.
\label{directions}
\end{equation}
Since the up quark mass matrix is diagonal to start with, the unitary matrix $V$ that diagonalizes $Y_d = V Y_d^D U$ on the left is the CKM matrix. However the quark bilinear $\bar{\mathcal{Q}}_i c_i\gamma_\mu \mathcal{Q}_i$, invariant under the transformation (\ref{U1cube}), reduces in the physical basis for the down quarks to
\begin{equation}
((c_3-c_1) \xi_{ij} + (c_2-c_1) V_{cj}V^*_{ci}) (\bar{d}_{Li}\gamma_\mu d_{Lj})
\end{equation}
which is not CKM-like unless $|c_2-c_1|\ll|c_3-c_1|$. For $|c_2-c_1| = O(1)$ this extra term would lead, from its effects on $\Delta S=2$ transitions, to a bound on the scale $\Lambda$ in the $10^3$ TeV range. 
Similar considerations hold for  the  case analogous to (\ref{U1cube}) in which the right handed up-quarks $u_i$ are interchanged with the the right handed down-quarks $d_i$, leading to a severe bound on $\Lambda$, in the hundreds of TeVs,  from $D^0$-$\bar{D}^0$ mixing.
Eqs.~(\ref{U1cube},\,\ref{directions}) and the case with $u_i$ reversed with $d_i$ are, respectively, examples of models of alignment in the up and down sector.
While alignment models can be successful in specific contexts, they appear not to work in the framework considered here.

\section{Status of MFV and \texorpdfstring{$U(2)^3$}{U(2)3} in meson mixings}\label{sec:DF2}

To determine the allowed room for new physics in meson mixings,
a global fit of the CKM matrix has to be performed.
In view of recent improved measurements, in particular the mixing induced CP asymmetry in $B_s\to J/\psi \phi$ and $B_s\to J/\psi f_0$ decays at LHCb that measures the $B_s$ mixing phase, as well as recent progress in the determination of relevant hadronic parameters from lattice QCD, we update the fits in our previous work \cite{Barbieri:2012uh} for MFV and $U(2)^3$.
In past years, the SM CKM fit has exhibited a sizable tension between the observables $|\epsilon_K|$, $S_{\psi K_S}=\sin(2\beta)$ and $\Delta M_s/\Delta M_d$ \cite{Lunghi:2008aa,Buras:2008nn,Barbieri:2011ci,Barbieri:2012uh} that could be solved in $U(2)^3$ models, but not in MFV.
To reassess the status of this tension,
we first perform a global CKM fit within the Standard Model.

\subsection{Standard Model CKM fit}

\begin{table}[tb]
\renewcommand{\arraystretch}{1.2}
 \begin{center}
\begin{tabular}{llllll}
\hline
$|V_{ud}|$ & $0.97425(22)$ &\cite{Aoki:2013ldr}& $f_K$  & $(156.3\pm0.8)$ MeV & \cite{Aoki:2013ldr}\\
$|V_{us}|$ & $0.2249(8)$ &\cite{Aoki:2013ldr}& $\hat B_K$ & $0.766\pm0.010$ &\cite{Aoki:2013ldr} \\
$|V_{cb}|$ & $(40.9\pm1.0)\times10^{-3}$ &\cite{Bevan:2013kaa}& $\kappa_\epsilon$ & $0.94\pm0.02$ & \cite{Buras:2010pza}\\
$|V_{ub}|$ & $(3.75\pm0.46)\times10^{-3}$ &\cite{Bevan:2013kaa}& $f_{B_s}\sqrt{\hat B_s}$  & $(262\pm10)$ MeV &\cite{Carrasco:2013zta}\\
$\gamma_{\rm CKM}$ & $(70.1\pm7.1)^\circ$ &\cite{Bevan:2013kaa}& $\xi$ & $1.225\pm0.031$ &\cite{Carrasco:2013zta}\\
$|\epsilon_K|$ & $(2.229\pm0.010)\times10^{-3}$ &\cite{Beringer:1900zz} &$\eta_{tt}$&$0.5765(65)$&\cite{Buras:1990fn}\\ 
$S_{\psi K_S}$ & $0.679\pm0.020$ &\cite{Amhis:2012bh} &$\eta_{ct}$&$0.496(47)$&\cite{Brod:2010mj}\\
$\Delta M_d$ & $(0.510\pm0.004)\,\text{ps}^{-1}$ &\cite{Amhis:2012bh} &$\eta_{cc}$&$1.87(76)$&\cite{Brod:2011ty}\\
$\Delta M_s/\Delta M_d$ & $(34.69\pm0.31)$ &\cite{Amhis:2012bh} &&&\\
$\phi_s$ & $0.01 \pm 0.07$ & \cite{Aaij:2013oba} &&&\\
\hline
 \end{tabular}
 \end{center}
\caption{Experimental inputs (left) and theoretical nuisance parameters (right) used in the $\Delta F=2$ fits.}
\label{tab:inputs}
\end{table}

We perform global fits of the CKM matrix using a Bayesian approach. We construct a likelihood from the experimental inputs listed in table~\ref{tab:inputs} and treat all the hadronic parameters listed there as nuisance parameters to be marginalized over.
From the full fit to all the relevant constraints, we obtain\footnote{The fitted value of $\lambda$ is given by the input value of $|V_{us}|$ to an excellent precision.}
\begin{align}
A &= 0.823\pm0.014 \,,
&
\bar\rho &= 0.142\pm0.020 \,,
&
\bar\eta &= 0.341\pm0.012
&
&\text{(full fit).}
\end{align}
To identify a possible tension between $S_{\psi K_S}$ and $\epsilon_K$, we perform two additional fits, leaving one of the two constraints out of the fit and thereby finding a ``fit prediction'' valid within the SM.
We find
\begin{align}
A &= 0.815\pm0.015 ,
&
\bar\rho &= 0.157\pm0.024 ,
&
\bar\eta &= 0.369\pm0.028 ,
&
S_{\psi K_S} &= 0.734\pm0.045
&
&\text{(no $S_{\psi K_S}$)}
\\
A &= 0.813\pm0.017 ,
&
\bar\rho &= 0.147\pm0.019 ,
&
\bar\eta &= 0.336\pm0.014 ,
&
\frac{|\epsilon_K|}{10^{-3}} &= (1.97\pm0.23)
&
&\text{(no $\epsilon_K$).}
\end{align}
Comparing to the experimental values in table~\ref{tab:inputs}, we find that the input value and fitted value for $S_{\psi \phi}$ as well as the input value and fitted value for $|\epsilon_K|$ both deviate by $1.1\sigma$.
We conclude 
that the data still prefer a negative contribution to the $B_d$ mixing phase $\phi_d = 2\beta + \phi_d^\Delta$ and/or a positive contribution to $|\epsilon_K|$, but the significance of this tension has decreased compared to the past. The main reasons for this change are the larger uncertainty in the QCD correction of the charm quark contribution to $\epsilon_K$ \cite{Brod:2011ty} as well as the significant upward shift in the central value of the kaon bag parameter from an average of lattice calculations \cite{Aoki:2013ldr}.
This conclusion is in line with recent fits by the UTfit collaboration \cite{UTfit} and is also in agreement with fits by the CKMfitter collaboration \cite{Charles:2013aka}.

\subsection{\texorpdfstring{$U(2)^3$ and MFV}{U(2)3} fits}\label{sec:u2fit}

In any new physics theory, contributions to the meson-antimeson mixing amplitudes in the $K$, $B_d$ and $B_s$ systems can be written as
\begin{align}
M_{12}^K &= (M_{12}^K)_\text{SM} \left( 1+ h_K e^{2i\sigma_K}\right) ,
&
M_{12}^{d,s} &= (M_{12}^{d,s})_\text{SM} \left( 1+ h_{d,s} e^{2i\sigma_{d,s}}\right).
\label{h-parameters}
\end{align}
Minimally broken $U(2)^3$ makes the prediction $\sigma_K=0$, $h_d=h_s\equiv h_B$, and $\sigma_d=\sigma_s\equiv \sigma_B$,
while MFV implies $\sigma_K=\sigma_d=\sigma_s=0$ and $h_d=h_s=h_K\equiv h$.
In both cases,
the correlation between $B_d$ and $B_s$ mixing implies
that the ratio of mass differences $\Delta M_s/\Delta M_d$, which has a smaller theory uncertainty than the individual mass differences, is not modified.
In the $U(2)^3$ case, one has in addition
a correlated shift in the mixing induced CP asymmetries $S_{\psi K_S}$ and $S_{\psi \phi}$,
while these observables are SM-like in MFV.
We have seen that the data prefers a positive $h_K$ as well as a positive contribution to the $B_d$ mixing phase, while a new physics contribution to the $B_s$ mixing phase is constrained already rather strongly.
To quantify the impact on the parameter space
of  $U(2)^3$ and MFV, we have performed two CKM fits, in the first case 
varying the $U(2)^3$ parameters $h_K$, $h_B$ and $\sigma_B$ in addition to the CKM Wolfenstein parameters and nuisance parameters and in the second case varying the single new parameter $h$ present in MFV.

\begin{figure}
\centering
\includegraphics[width=0.7\textwidth]{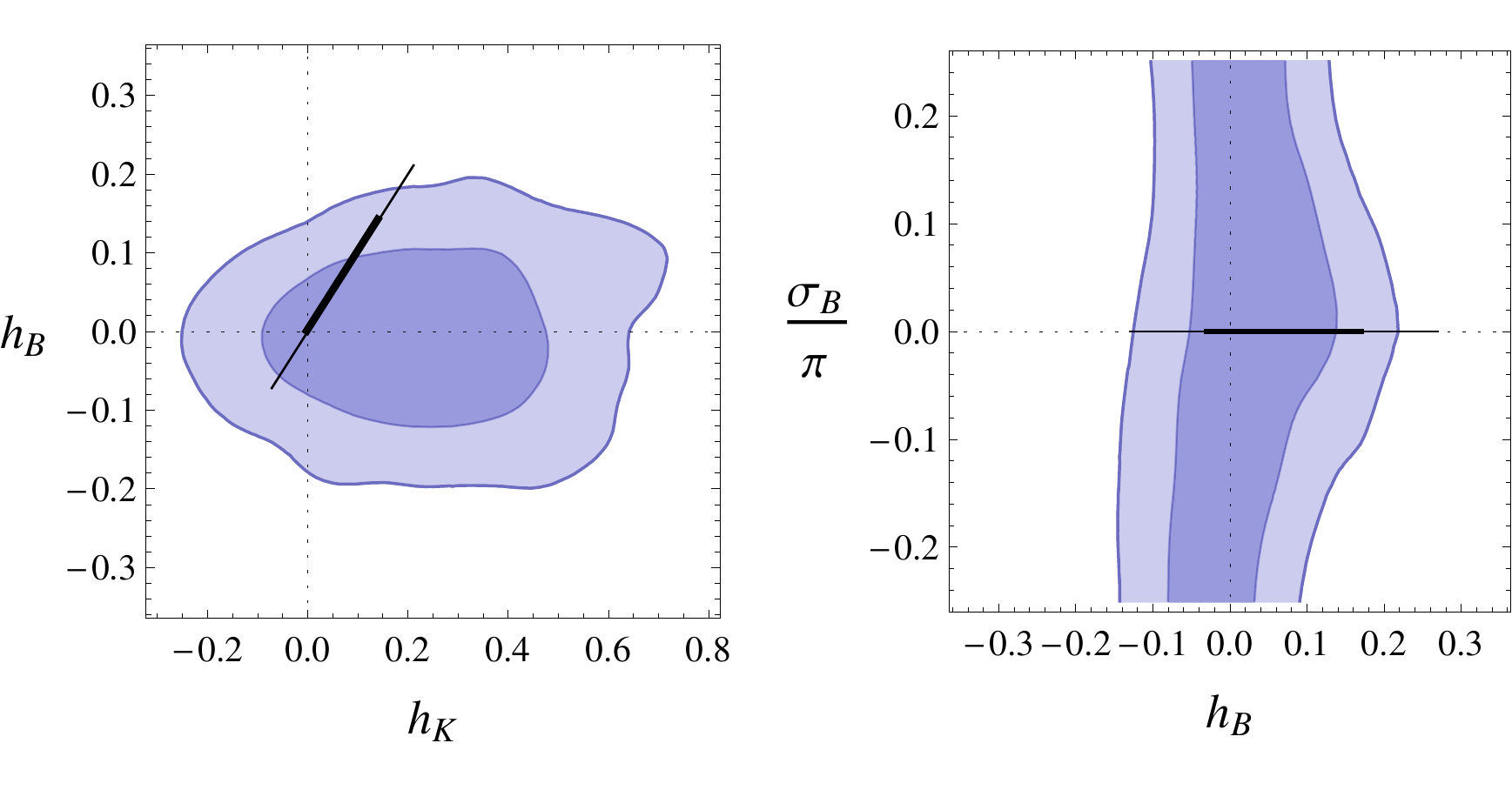}
\caption{Results of the $U(2)^3$ CKM fit. The dark and light regions correspond to 68\% and 95\% probability, respectively. The segments show the corresponding results for MFV, the thick and thin lines corresponding to 68\% and 95\% probability.}
\label{fig:DF2plots1}
\end{figure}

The fit results for $U(2)^3$ are shown in fig.~\ref{fig:DF2plots1}. It shows a preference for $h_K>0$ caused by the $\epsilon_K$ tension, while there is no clear preference for a non-zero value of the CP-violating phase $\sigma_B$. This is due to the limited allowed room for new physics in the phase $\phi_s$, which precludes a sizable negative contribution to the $B_d$ mixing phase, that would otherwise be preferred.
The fit result in the MFV case is $h=0.07\pm0.10$ and is shown in the same figure as a black line.
Here, the preference for a positive $h$ is much milder, since a positive contribution to $\epsilon_K$ would also imply a positive contribution to $\Delta M_d$, which is not preferred by the data.

\begin{figure}
\centering
\includegraphics[width=0.7\textwidth]{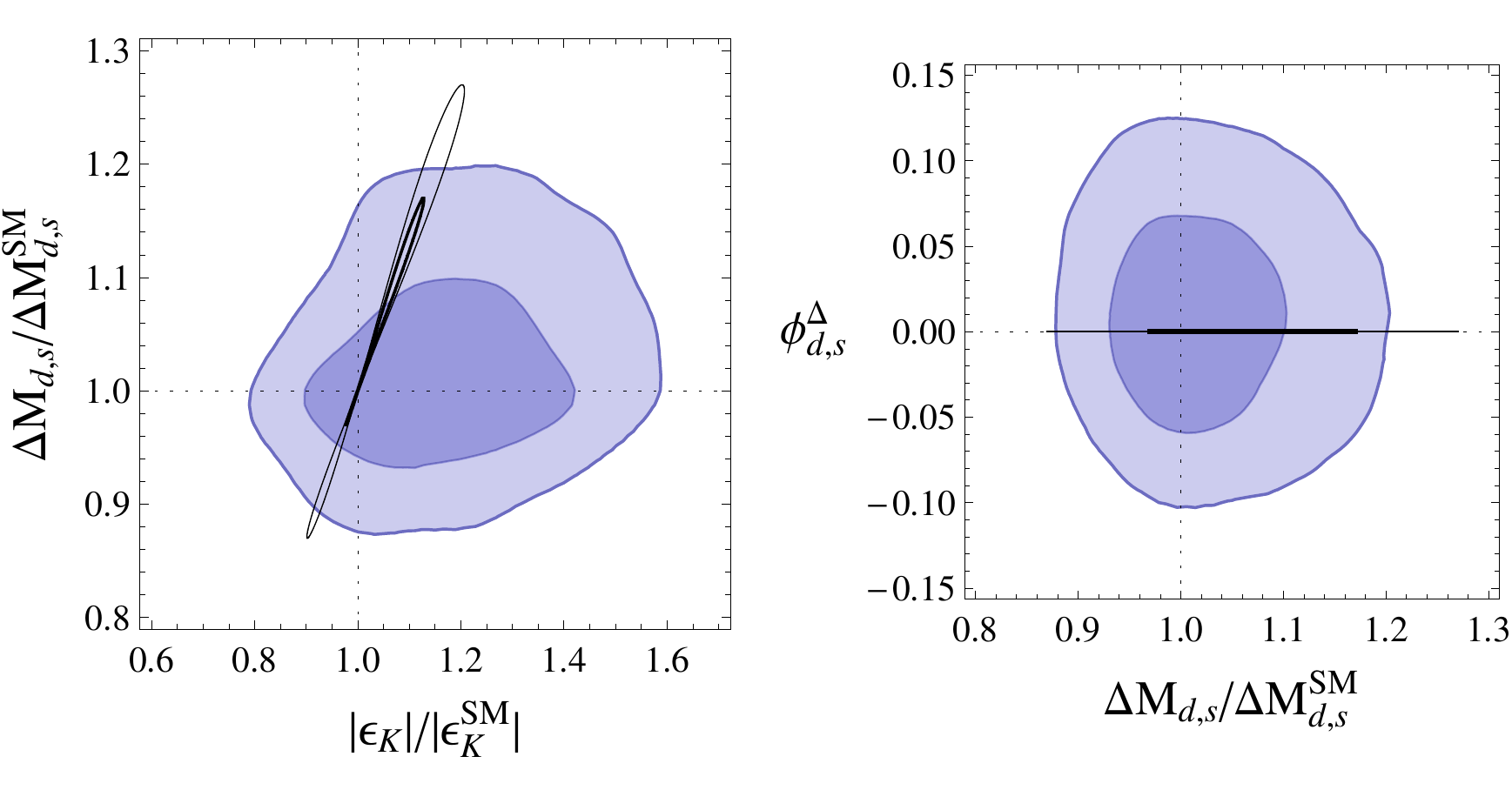}
\caption{Results of the $U(2)^3$ CKM fit. The dark and light regions correspond to 68\% and 95\% probability, respectively.
The black regions show the corresponding results for MFV, the thick and thin lines corresponding to 68\% and 95\% probability.
}
\label{fig:DF2plots2}
\end{figure}

The parameters $h_K$, $h_B$ and $\sigma_B$ can also be related to measurable quantities by means of the relations
\begin{align}
\frac{\Delta M_{d,s}}{\Delta M_{d,s}^\text{SM}}&=\left|1+h_B e^{2i\sigma_B}\right|
\,,&
\phi_{s,d}^\Delta&=\text{arg}\left(1+h_B e^{2i\sigma_B}\right)
\,,&
\frac{|\epsilon_K|}{|\epsilon_K^\text{SM}|}&\approx 1+ \frac{|\epsilon_K^{tt,\text{SM}}|}{|\epsilon_K^\text{SM}|} \, h_K 
\label{eq:h2obs}
\end{align}
where we used the conventions
\begin{align}
 \phi_s&=-2|\beta_s|+\phi_s^\Delta\,, &S_{\psi K_S}&= \sin(2\beta + \phi_d^\Delta)\,.
\end{align}
The best-fit results for these observable quantities are shown in fig.~\ref{fig:DF2plots2}, again showing the result for the MFV fit as black lines\footnote{In the left plot, the line spreads to a narrow region due to the uncertainty on the ratio  ${|\epsilon_K^{tt,\text{SM}}|}/{|\epsilon_K^\text{SM}|}$ in eq.~(\ref{eq:h2obs}).}.

The correspondence between the phenomenological $h$ parameters in eq. (\ref{h-parameters}) and the coefficients $C$ in eqs. (\ref{CLL}) and (\ref{eff1}, \ref{eff2}) is given by
\begin{equation}
h\approx C \left[\frac{3.1~\text{TeV}}{\Lambda}\right]^2, 
\label{h_Lambda}
\end{equation}
so that, for $C=1$, the current measurements are probing scales of about 4 to 7~TeV.

With similar coefficients, the $\Delta F=1$ operators in section~\ref{eff-ops} are at present generally less constrained than the $\Delta F=2$ operators discussed in this section\footnote{A possible exception is the $\epsilon^\prime/\epsilon$ constraint on the operators $(\bar{s}_L \gamma_\mu d_L)(\bar{q}_R \gamma_\mu q_R)$ and $(\bar{s}_L^\alpha \gamma_\mu d_L^\beta)(\bar{q}_R^\beta \gamma_\mu q_R^\alpha)$ \cite{Barbieri:2012bh}. In the case of supersymmetry that we will analyse, however, all contributions to such operators are further suppressed by the heaviness of the first generation squarks and/or by powers of small Yukawa couplings.}. We shall nevertheless elaborate on them,  making this statement more precise, in the specific case of supersymmetry, dealt with in the rest of the paper.

\section{Status of flavour physics in SUSY \texorpdfstring{$U(2)^3$}{U(2)³}}\label{sec:DF2-SUSY}

As seen in the previous section, in minimally broken $U(2)^3$ models the data still allow for a $20\%$ enhancement of $\Delta M_{d,s}$, a $60\%$ enhancement of $\epsilon_K$ and a contribution to the $B_{d,s}$ mixing phases of $\pm0.1$\,rad. In supersymmetry, whether such values can be reached -- i.e.\ whether $\Delta F=2$ observables still present a relevant constraint -- depends crucially on the direct bounds on sparticle masses set by the LHC experiments. In this section, we first carefully derive all the relevant SUSY contributions to meson mixing, then discuss the LHC bounds on sparticle masses, and finally perform a numerical analysis of the allowed $\Delta F=2$ effects in SUSY $U(2)^3$. The supersymmetric spectrum that we have in mind is motivated by naturalness with, as its main feature, first and second generation squarks heavier than all other supersymmetric particles.

\subsection{Anatomy of SUSY contributions to meson mixing}\label{sec:DF2-SUSY-ana}

The dominant contributions to $\Delta F=2$ observables in SUSY $U(2)^3$ come from gluino, Wino, Higgsino and charged Higgs diagrams\footnote{Bino and mixed gluino-neutralino diagrams are typically subleading, but will be taken into account in the numerical analysis of section~\ref{sec:DF2-SUSY-numerics}.}. Previously \cite{Barbieri:2011ci},
% ,Blankenburg:2012ah},
only gluino contributions were considered for simplicity but this approximation becomes questionable in view of the stringent LHC bounds on the gluino mass. Instead, we discuss all the relevant contributions in turn.

\paragraph{Charged Higgs and Higgsino contributions}

\begin{figure}
\centering%
\includegraphics[width=.47\textwidth]{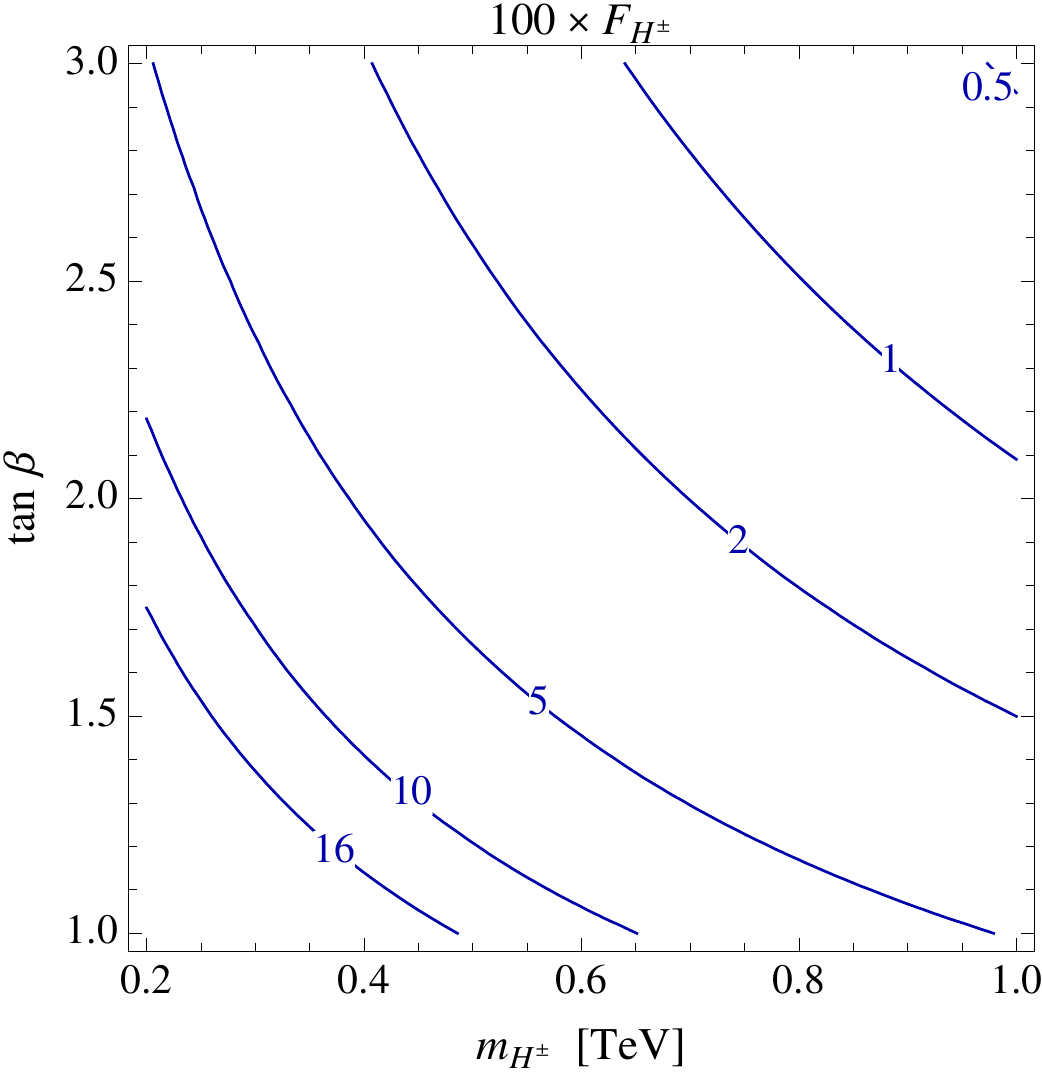}\hfill%
\includegraphics[width=.47\textwidth]{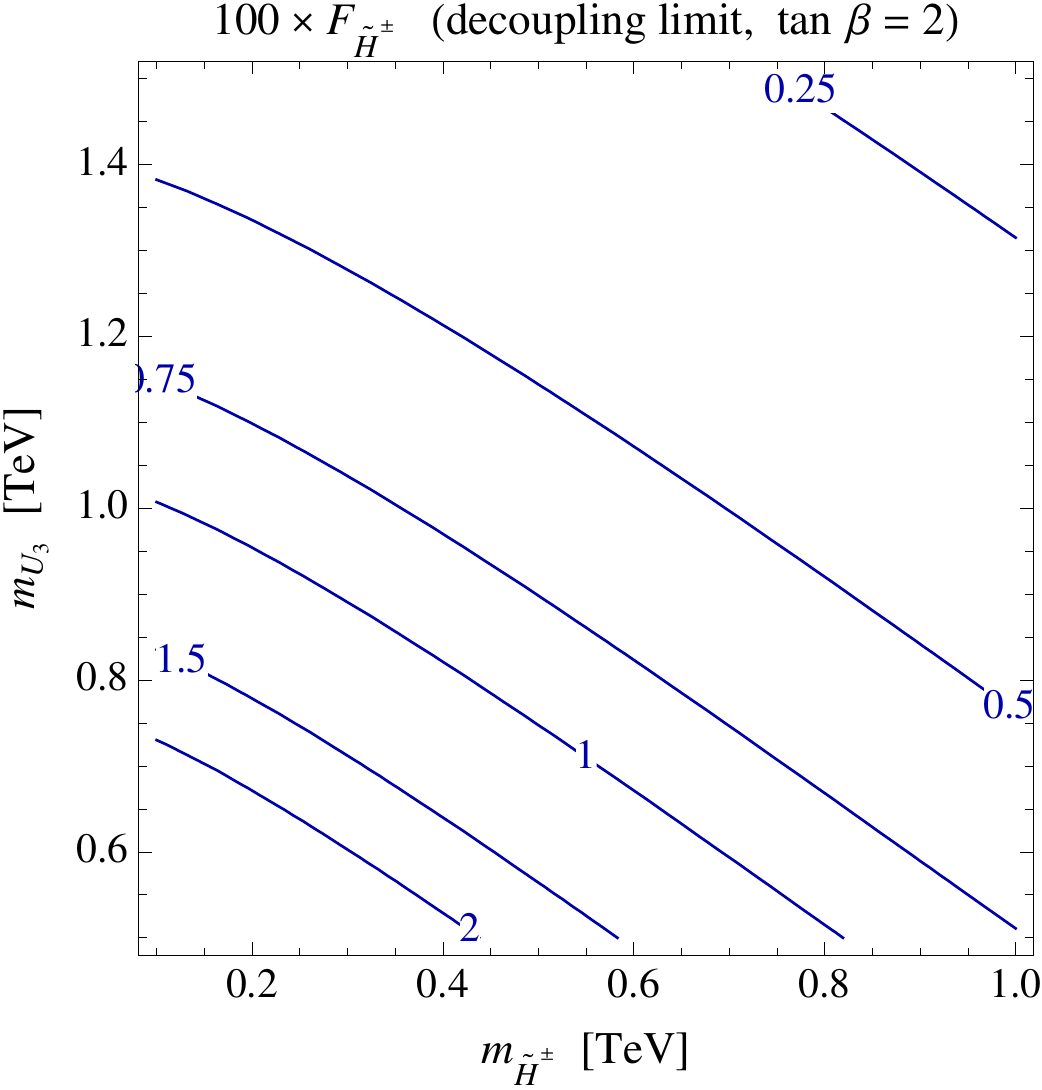}
\caption{Charged Higgs and Higgsino contributions to $\Delta F = 2$ observables.\label{higgsino}}
\end{figure}

Being proportional to the Yukawa couplings, they enter the $K$, $B_d$ and $B_s$ mixing amplitudes universally and aligned in phase with the SM, just as in the case of $U(3)^3$ (MFV). One can write
\begin{align}
[h_K]_{H^\pm, \tilde H^\pm} = [h_B]_{H^\pm, \tilde H^\pm} \equiv  F_{H^\pm}+ F_{\tilde H^\pm}\,.
\label{eq:FchargedH}
\end{align}
The functions $F_{H^\pm}$ and $F_{\tilde H^\pm}$ are shown in fig.~\ref{higgsino} as functions of the charged Higgs or Higgsino and right-handed stop masses, $m_{H^\pm}$, $m_{\tilde H^\pm}$ and $m_{U_3}$.

\paragraph{Gluino contributions}

\begin{figure}
\centering%
\includegraphics[width=.47\textwidth]{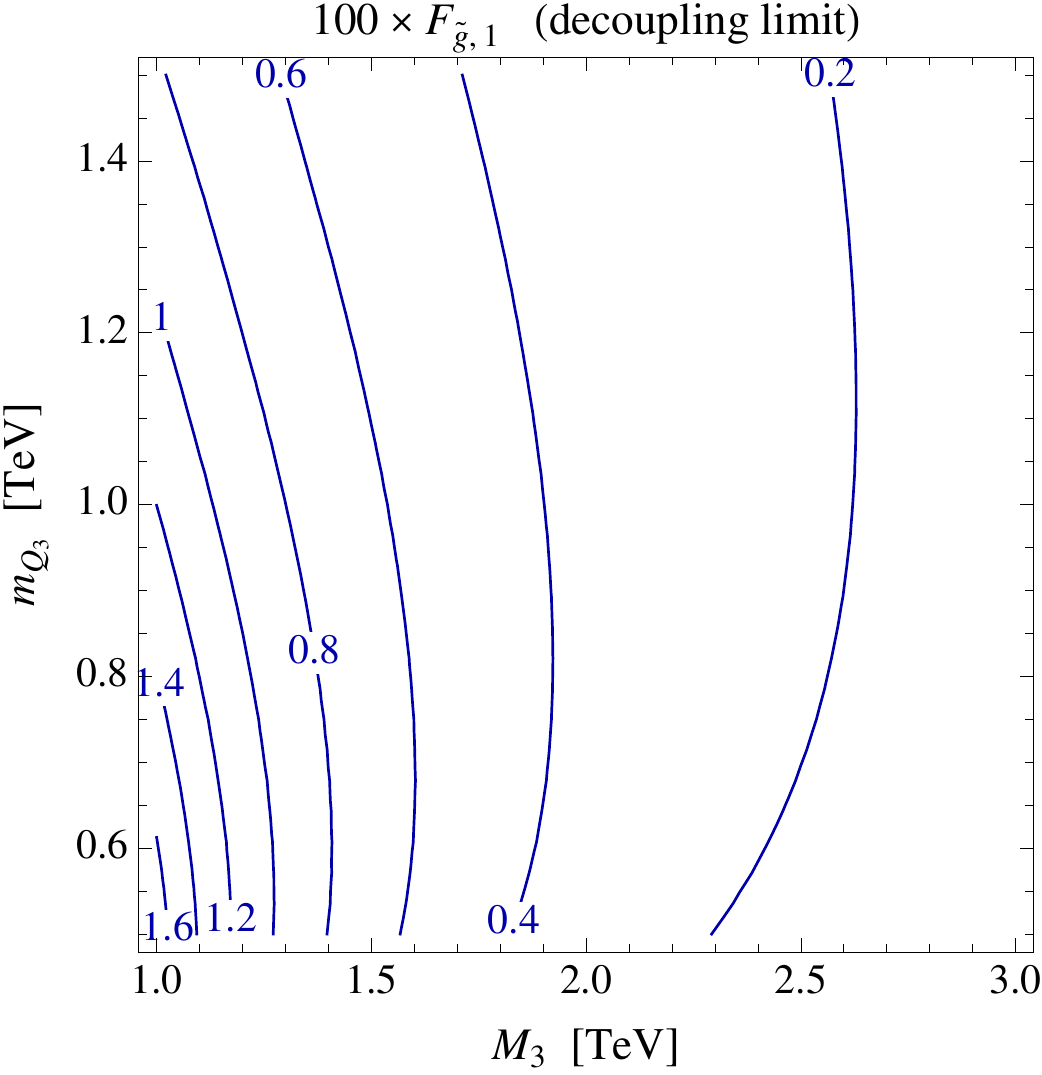}\hfill%
\includegraphics[width=.47\textwidth]{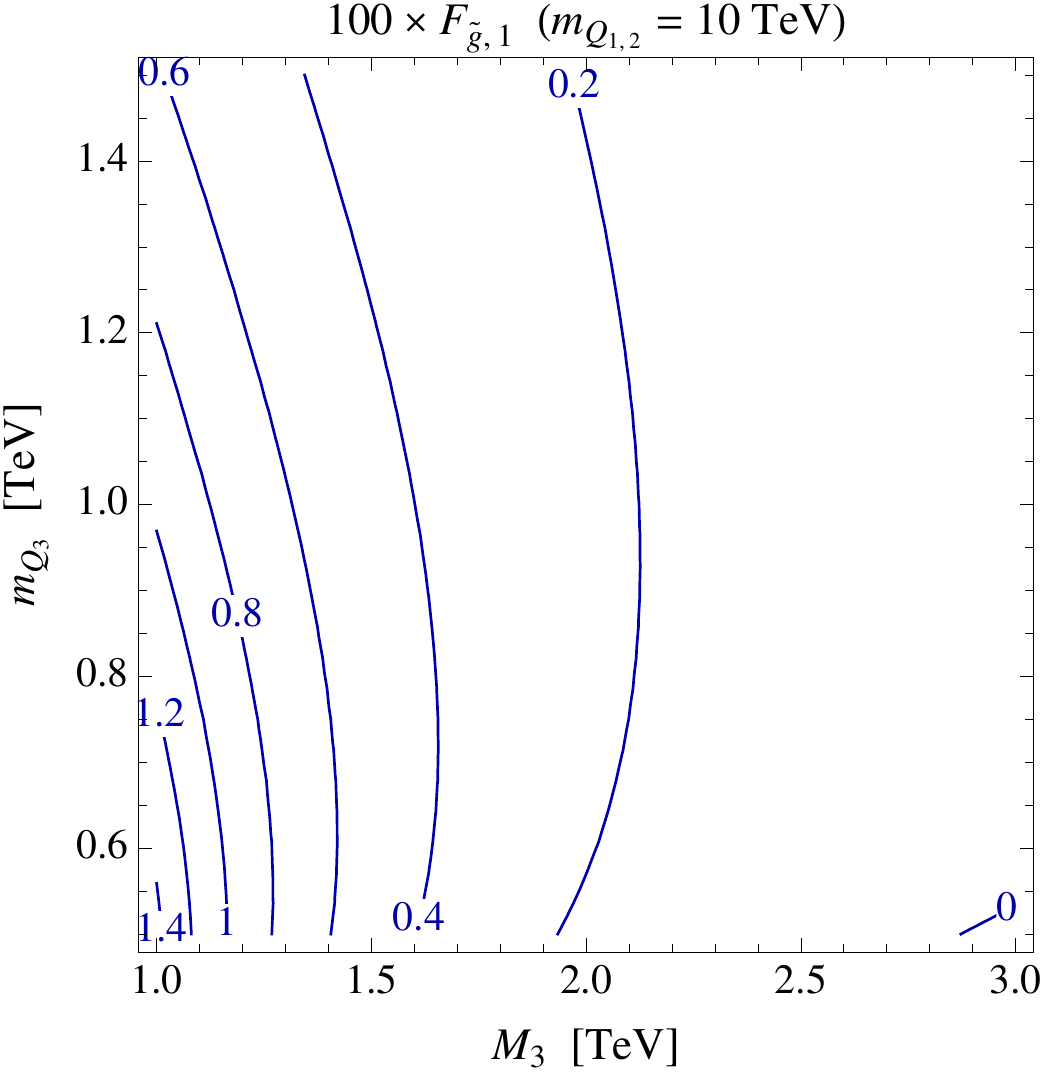}
\caption{``33'' gluino contribution to meson mixings in the decoupling limit (left), and with degenerate first two generation squarks at 10 TeV (right).\label{glu1}}
\end{figure}

\begin{figure}
\centering%
\includegraphics[width=.47\textwidth]{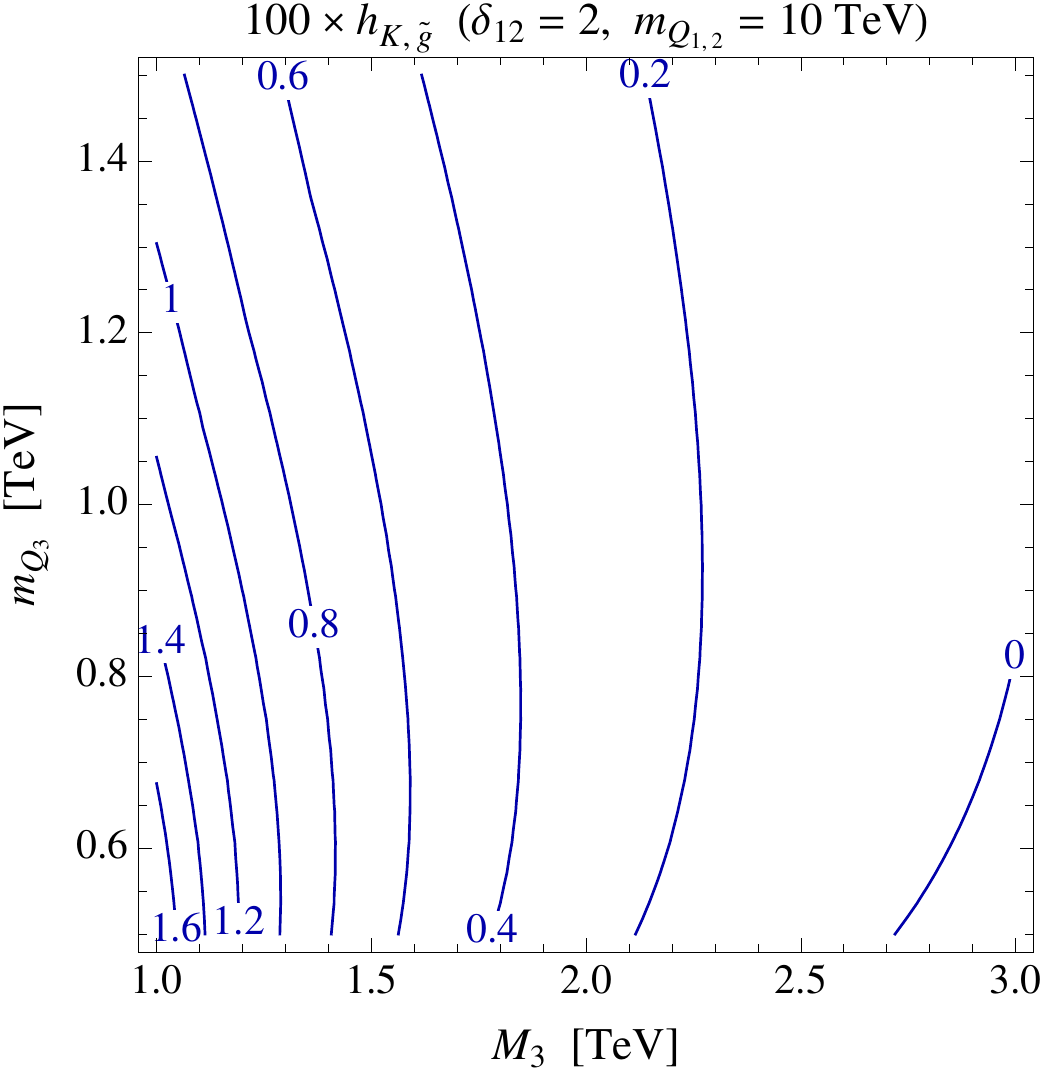}\hfill%
\includegraphics[width=.47\textwidth]{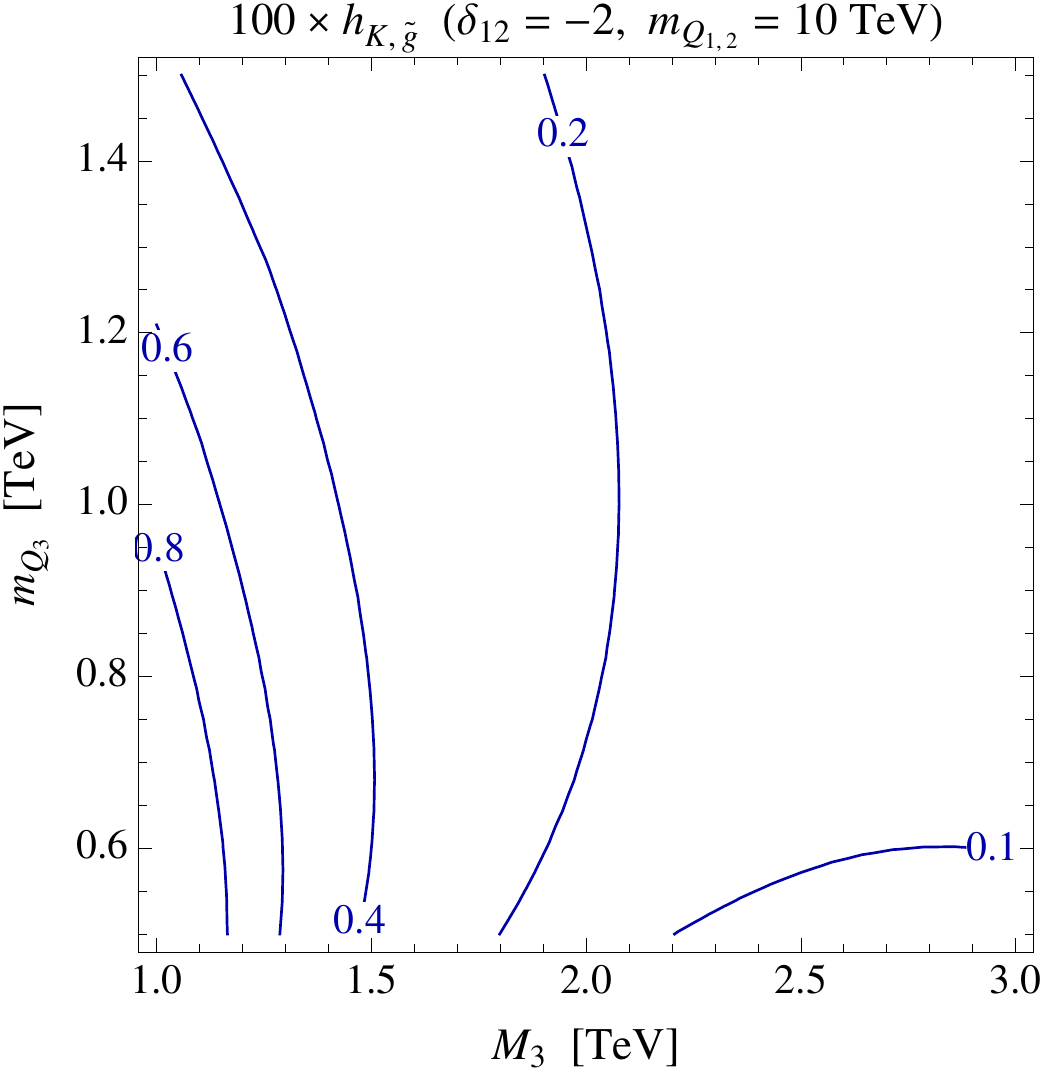}
\caption{Effect of the splitting between the first two generation squarks on the gluino contribution to $h_K$. Left: $\delta_{12} = 2$ ($m_{Q_1} > m_{Q_2}$). Right: $\delta_{12} = -2$ ($m_{Q_1} < m_{Q_2}$). The average heavy mass is 10 TeV and $\xi_L = 1$.\label{glu2}}
\end{figure}

\begin{figure}
\centering%
\includegraphics[width=.47\textwidth]{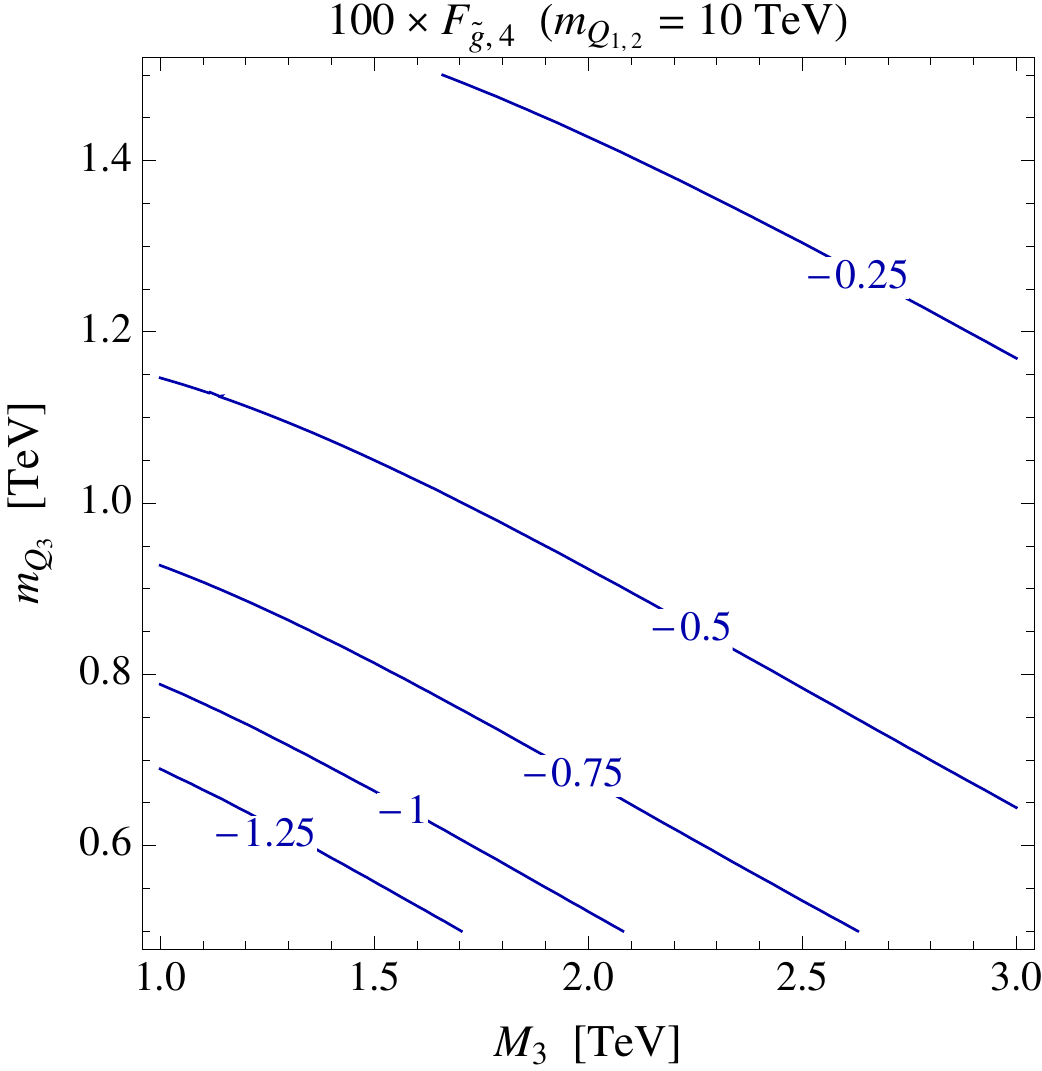}\hfill%
\includegraphics[width=.47\textwidth]{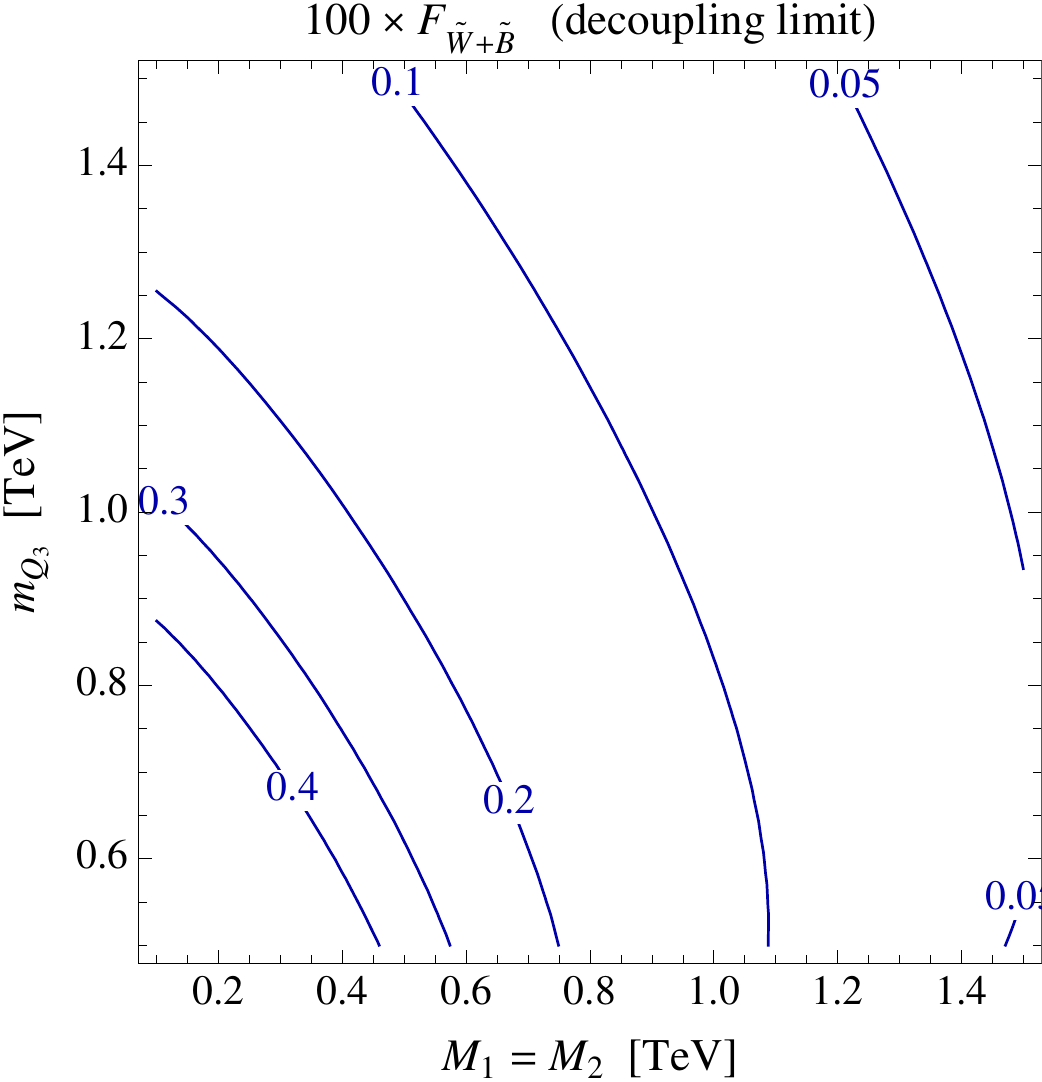}
\caption{Left: Contribution of the LR gluino operator to $B_s$ mixing; the first two generation squark masses are set to 10 TeV. Right: Wino contribution to meson mixings in the decoupling limit for the first two generations of squarks.\label{glu3}}
\end{figure}

For the discussion of the relevant effects, it is useful to group the contributions with different squark flavours into ``33'', ``12'', and ``12,3'' contributions \cite{Barbieri:2010ar}. In the decoupling limit for the first two generation squarks considered in \cite{Barbieri:2011ci}, the ``12'' and ``12,3'' contributions vanish and the ``33'' contribution only depends on the left-handed sbottom mass $m_{Q_3}$.
Beyond this limit, there are three potentially relevant additional effects,
\begin{itemize}
\item for first and second generation squarks in the few TeV range, the ``33'' contribution has a non-negligible dependence on the first two generation squark masses (see fig.~\ref{glu1}), in particular when the gluino mass $M_3\gg m_{Q_3}$ since there is a numerical cancellation in the loop function of the leading contribution;
\item since the first two generation squark masses are not exactly degenerate, there is a non-zero effect in the ``12'' and ``12,3'' contributions (see fig.~\ref{glu2}) that depends on an additional free $O(1)$ parameter and is numerically relevant for $K$ mixing;
\item although right-handed flavour violation is strongly suppressed in minimally broken $U(2)^3$, it turns out that the contribution to the LR operator is not negligible in $B_s$ mixing (see fig.~\ref{glu3} left). We explicitly checked that all the non-gluino LR contributions are instead negligible.
\end{itemize}
The first and third points have also been discussed in \cite{Blankenburg:2012ah}.
The gluino contributions to the meson mixing amplitudes can thus be written as
\begin{align}
[h_K]_{\tilde g} &=
|\xi_L|^4 F_{\tilde g,1} + |\xi_L|^2\delta_{12} F_{\tilde g,2} + |\delta_{12}|^2 F_{\tilde g,3}
\,, \label{eq:gluino_K}\\
[h_d \, e^{2i\sigma_d}]_{\tilde g} &=
|\xi_L|^2 \,e^{2i\gamma_L} F_{\tilde g,1}
\,,\label{eq:gluino_Bd}\\
[h_s \, e^{2i\sigma_s}]_{\tilde g} &=
|\xi_L|^2 \,e^{2i\gamma_L} F_{\tilde g,1} + |\xi_L\xi_R| \,e^{i(\gamma_L+\gamma_R)} F_{\tilde g,4} 
\,,\label{eq:gluino_Bs}
\end{align}
where $\xi_{L,R}$ and $\delta_{12}$ are free $O(1)$ parameters and $\gamma_{L,R}$ are CP-violating phases \cite{Barbieri:2011ci,Barbieri:2011fc}. Note that the parameters $\xi_R$ and $\delta_{12}$ control effects that are of second order in the breaking parameters in eq. (\ref{U2^3}).
Explicit analytic expressions for the loop-functions $F_{\tilde g,i}$ are given in the Appendix.

\paragraph{Wino contributions}

They are analogous to the gluino contributions discussed above. While they are parametrically suppressed by smaller gauge couplings compared to the gluino contributions, this can be compensated in practice by the much weaker bound on the Wino mass compared to the gluino mass. Bino and mixed Wino-Bino contributions are analogous, although further suppressed by powers of $\sin\theta_w$ and thus negligible.

\subsection{LHC mass bounds}\label{sec:direct}

To address the question of how large the effects in flavour physics can still be after the first LHC phase, we need to impose the bounds on sparticle masses resulting from various LHC searches. While a full analysis of this question is  beyond the scope of the present study, experimental analyses of simplified models provide a rough indication of the exclusion ranges \cite{TheATLAScollaboration:2013mha,TheATLAScollaboration:2013tha,ATLAS:2013tma,Chatrchyan:2013wxa,Chatrchyan:2013iqa,Chatrchyan:2013fea,CMS:2013cfa,CMS:2013ida,CMS:2013ija,ATLAS:2013cma,ATLAS:2013pla,TheATLAScollaboration:2013xha, TheATLAScollaboration:2013gha, TheATLAScollaboration:2013aia,Aad:2013ija,Chatrchyan:2013xna,Chatrchyan:2013lya,ATLAS:2013rla,TheATLAScollaboration:2013hha,ATLAS:2013yla,TheATLAScollaboration:2013zia,CMS:2013dea,Chatrchyan:2012paa,TheATLAScollaboration:2013fha}. Throughout, we assume a neutralino LSP, decoupled first and second generations squarks, and decoupled sleptons.

In the case of the gluino, searches for gluino pair production with four top quarks in the final state rule out a gluino below $1.4$~TeV for LSP masses below 700~GeV if the mass difference between the gluino and the LSP is larger than $2m_t$ \cite{TheATLAScollaboration:2013mha,TheATLAScollaboration:2013tha,ATLAS:2013tma,Chatrchyan:2013wxa,Chatrchyan:2013iqa,Chatrchyan:2013fea,CMS:2013cfa,CMS:2013ida,CMS:2013ija}. If the mass difference is smaller than that, searches for events with light quark jets and missing energy rule out an LSP below 400~GeV \cite{TheATLAScollaboration:2013fha}.

Stops below about 700~GeV are ruled out for LSP masses below 200 GeV, unless the mass difference between the stop and LSP is below 200 GeV; in that case, there is still some room for a light stop \cite{ATLAS:2013cma,ATLAS:2013pla,TheATLAScollaboration:2013xha, TheATLAScollaboration:2013gha, TheATLAScollaboration:2013aia,Aad:2013ija,Chatrchyan:2013xna,CMS:2013cfa}.
Similarly, sbottom squarks are excluded below 650~GeV for an LSP below 250~GeV \cite{Aad:2013ija,Chatrchyan:2013lya,CMS:2013ida}.

We find no relevant bound on charginos and neutralinos, among those of \cite{ATLAS:2013rla,TheATLAScollaboration:2013hha,ATLAS:2013yla,TheATLAScollaboration:2013zia,CMS:2013dea,Chatrchyan:2012paa}, given our assumptions of heavy sleptons and a neutralino LSP. 
Notice also that we ignore the exclusion in the $m_A$-$\tan \beta$ plane from negative Higgs boson searches at LEP \cite{Schael:2006cr}. Despite being specific to the MSSM, it was cast for $M_{\rm SUSY} = 1$ TeV, and in the region of low $\tan \beta$ where it is relevant a larger value of $M_{\rm SUSY}$ is required to reproduce the measured value of the Higgs mass \cite{Djouadi:2013vqa}.

\subsection{Numerical analysis of meson mixing}\label{sec:DF2-SUSY-numerics}

Having identified the relevant contributions and their typical size, in this section we perform a numerical scan of the SUSY $U(2)^3$ parameter space to identify the allowed effects in the $\Delta F=2$ observables.
To this end, we scanned the relevant parameters in the following ranges,
\begin{align}
m_{Q_3}, m_{U_3}, m_{D_3} &\in [0.1,1.5] \,\text {TeV}\,,&
M_{3} &\in [0.1,3] \,\text {TeV}\,, \label{scan_glustop}\\
m_{Q_{1,2}}, m_{U_{1,2}}, m_{D_{1,2}}%,m_L,m_E
&\in [10,30] \,\text {TeV}\,,&
|\mu| &\in [0.1,0.75] \,\text {TeV}\,, \label{scan_firstmu}\\
|M_1|,|M_2|,M_A &\in [0.1,0.8] \,\text {TeV}\,,&
A_{u,t}/m_{Q_3,U_3} &\in [-1,1]\,, \label{scan_EWAu} \\
A_{d,b}/m_{Q_3,D_3} &\in [-1,1] \,,&
\tan\beta &\in [1,5] \,, \label{scan_Adtbeta}  \\
|\text{all $O(1)$ parameters}| &\in [1/3,3]\,. \label{scan_O1}
\end{align}
We used log priors for the masses and linear priors for the dimensionless parameters. We scanned the phases of the $O(1)$ parameters that are allowed to be complex, but assumed $\mu,M_1,M_2$, and the trilinears to be real.
This is motivated by the recent result on the electron electric dipole moment and the considerations developed in section~\ref{EDM}.

After computing the mass spectrum, we imposed the LHC direct bounds in a simplified manner as discussed in section~\ref{sec:direct}.
The points that survive after imposing these direct constraints either have stop and sbottom masses beyond 650~GeV and gluinos beyond 1.3~TeV, or they have a partially compressed spectrum with a small mass difference between the gluino and LSP or the stop and LSP.
In the scatter plots presented below, we distinguished these two classes of spectra by dark blue and light blue points, respectively.

Finally, we used {\tt SUSY\_FLAVOR 2.10} \cite{Crivellin:2012jv} to compute flavour observables. To provide a  check of the numerical results we have constructed an independent program, finding very good agreement. We imposed the constraint from the branching ratio of $B\to X_s\gamma$ at $2\sigma$ (the points not passing this constraint are shown in light gray in the plots). The $B_s\to\mu^+\mu^-$ constraint is irrelevant at the low values of $\tan\beta$ we are considering.
The range $\tan{\beta} =1\div 5$ is chosen because it makes easier to accommodate in the Next to Minimal Supersymmetric Standard Model a 125~GeV Higgs boson with stops below 1~TeV and because of the $B\to X_s\gamma$ and electron EDM constraints discussed in the following sections.

\begin{figure}[tbp]
\centering
\includegraphics[width=0.79\textwidth]{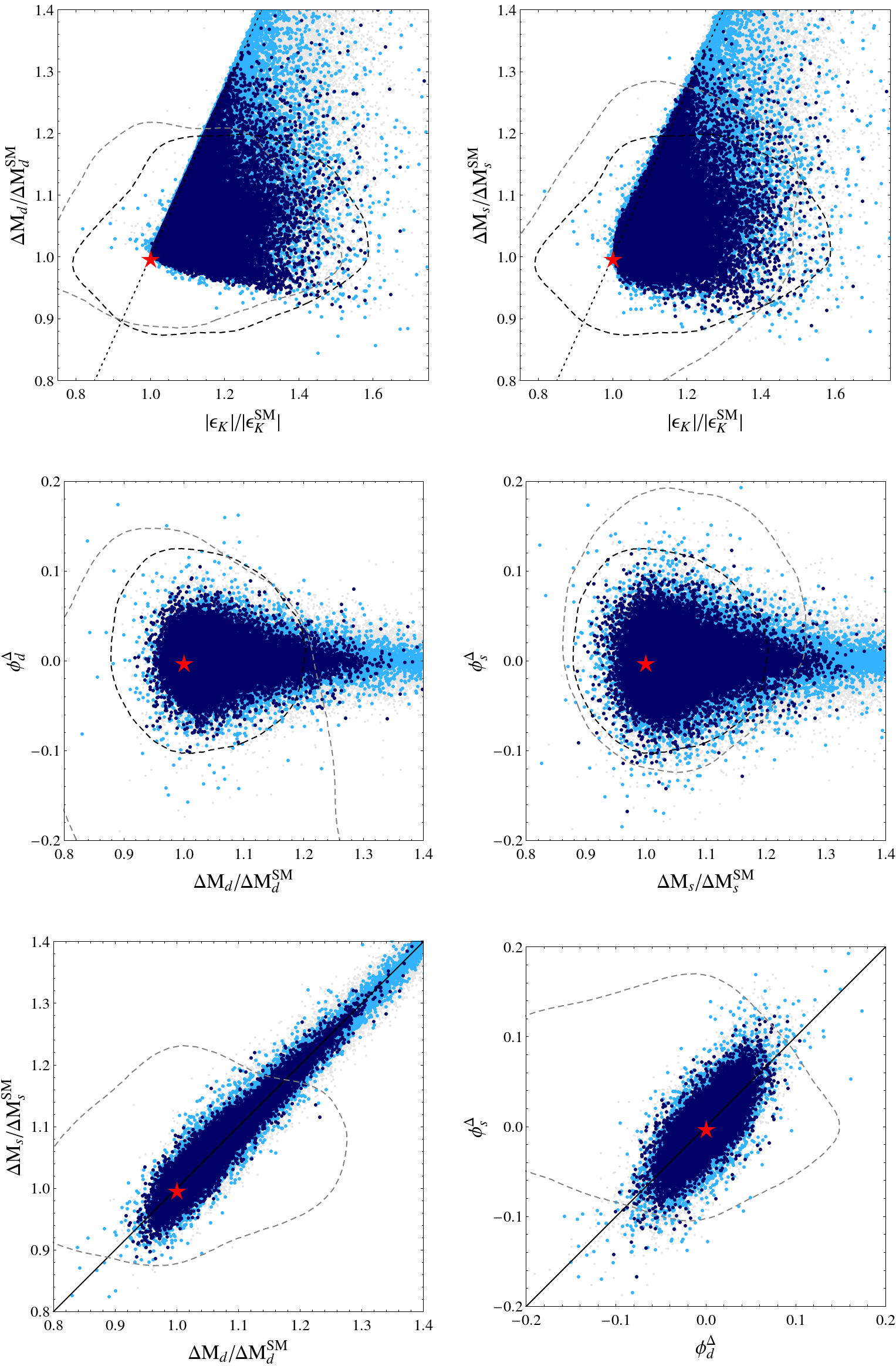}
\caption{Correlation between $\Delta F=2$ observables. Dark and light blue points are allowed by all constraints, light blue points have a compressed spectrum (see text for details), light gray points are ruled out by $B\to X_s\gamma$. The red star is the SM. The black dashed line is the 95\% C.L.\ region allowed by the global CKM fit imposing the $U(2)^3$ relations, the gray dashed line in a generic new physics fit. The dashed lines in the first row show the MFV limit, the solid lines in the last row the $U(2)^3$ limit.}
\label{fig:scatter}
\end{figure}

The predictions for the $\Delta F=2$ observables are shown in fig.~\ref{fig:scatter}.
In all the plots, the red star shows the SM value.
The dotted black line in the first two upper plots corresponds to the $U(3)^3$ limit of a CKM-like phase in $B_{d,s}$ and $K$ mixing. As discussed in sec.~\ref{sec:DF2-SUSY-ana}, this limit is adhered to by the Higgs and Higgsino contributions.
In the two lower plots, the solid black line corresponds to the $U(2)^3$ limit of universality in $b\to d,s$ transitions.
We stress that the violation of this limit visible in the plots is not due to the uncertainty in CKM parameters -- since we plot only the ratio of mass differences and the new physics contribution to the mixing phase, these uncertainties cancel out -- but it is due to the contribution of the LR operator to $B_s$ mixing
discussed in sec.~\ref{sec:DF2-SUSY-ana}\footnote{Notice that this contribution also spoils the $V_{ub}$--$S_{\psi\phi}$--$S_{\psi K_S}$ correlation pointed out in \cite{Buras:2012sd}.}.
In all the plots, the dashed black contour is the 95\% probability region allowed in the $U(2)^3$ CKM fit of section~\ref{sec:u2fit} (cf.\ fig.~\ref{fig:DF2plots2}).
For comparison, we also show as gray dashed contours the allowed regions in a more general fit not requiring universality of $B_d$ and $B_s$ mixing. In this case, a sizably negative value of $\phi_d^\Delta$ is allowed because of the $\sin(2\beta)$ tension.

\subsection{Predictions for rare $B$ decays}

Gluino contributions to rare $B$ decay amplitudes have been discussed by us in ref.~\cite{Barbieri:2011fc}, were we showed that the main new effects in $U(2)^3$ arise from the magnetic and chromomagnetic dipole operators. In addition to the gluino (and Wino) contributions, which are proportional to $e^{i \gamma_L}$ and hence contribute to CP asymmetries in $B$ decays, there are the usual MFV-like Higgsino and charged Higgs contributions to dipole operators, that are aligned in phase with the SM contribution and are mainly constrained by the branching ratio of $B\to X_s\gamma$.
The most promising observable to probe the CP-violating contributions is the angular CP asymmetry $A_7$ in the $B\to K^*\mu^+\mu^-$ decay at low dimuon invariant mass $q^2$ \cite{Bobeth:2008ij,Altmannshofer:2008dz}.
Correlated with this, contributions to the mixing-induced CP asymmetries in non-leptonic penguin decays like $B\to \phi K_S$ or $B\to \eta' K_S$ will be generated by the CP violating contribution to the chromomagnetic dipole operator \cite{Barbieri:2011fc}.

\begin{figure}[tbp]
\centering
\includegraphics[width=0.5\textwidth]{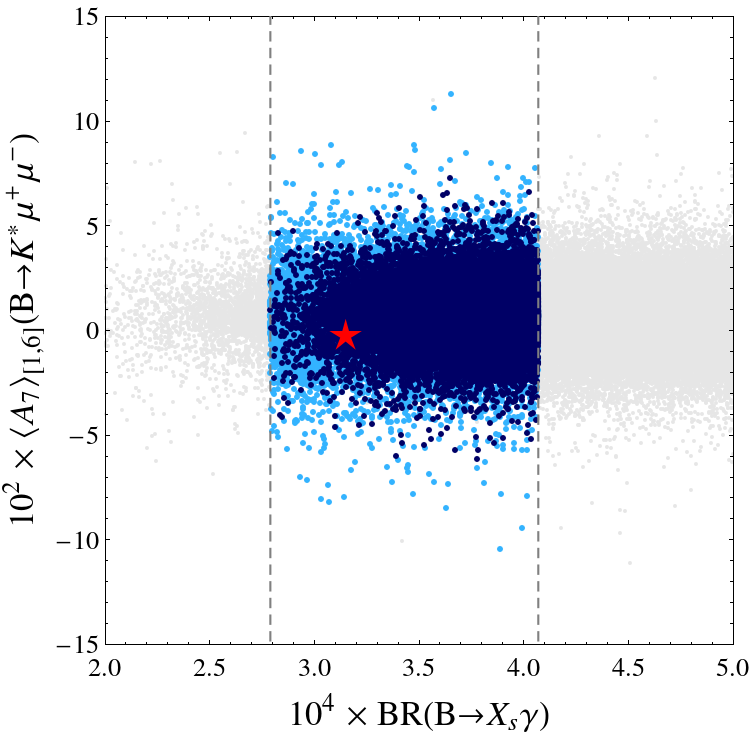}
\caption{Branching ratio of $B\to X_s\gamma$ vs.\ the angular CP asymmetry $A_7$ in $B\to K^*\mu^+\mu^-$, integrated from $q^2=1$ to 6~GeV$^2$.}
\label{fig:bsg}
\end{figure}

Fig.~\ref{fig:bsg}
shows numerical results for the branching ratio of $B\to X_s\gamma$ vs.\ the CP asymmetry $A_7$ integrated in $q^2$ from 1 to 6~GeV$^2$.
The scan ranges and the colours are as in sec.~\ref{sec:DF2-SUSY-numerics}.
While the MFV-like contributions can lead to large modifications of the branching ratio that are constrained experimentally, the effects in $A_7$ are limited due to the strong direct bounds, in particular on the gluino mass.

We note here that for $\Delta B=1$ processes, the parameter $\tan\beta$ plays a more important role than for $\Delta B=2$ processes. First, this is because the dipole operators change chirality and hence in general they can receive $\tan\beta$ enhanced contributions. Second, there are the $B_{s,d}\to\mu^+\mu^-$ decays which receive strongly $\tan\beta$ enhanced contributions from scalar operators.
Here, we focus on the regime $\tan\beta\lesssim5$, where the branching ratios of $B_{s,d}\to\mu^+\mu^-$ are modified by at most 30\% with respect to the SM.

\subsection{Electron electric dipole moment}
\label{EDM}

Suppressing SUSY contributions to Electric Dipole Moments  (EDM) is an additional motivation for a split squark spectrum, as it allows to have sizable CP-violating phases without excessive one-loop contributions to EDMs of first generation fermions.
Recently, a new experimental bound on the electron EDM has been obtained \cite{Baron:2013eja},
$|d_e| < 8.7\times 10^{-29}\,e\,\text{cm}$,
that improves the previous bound \cite{Hudson:2011zz} by a factor of 12.
Here, we study the impact of this new bound on models with a split sfermion spectrum.

First, there is the direct one-loop contribution to the electron EDM involving charginos and sneutrinos, that decouples with the scale of the first generation sfermion masses. Updating the bound in \cite{Barbieri:2011vn},
we find a lower bound on the sneutrino mass, depending on $\tan\beta$ and the phase of the $\mu$ term ($\mu=|\mu|e^{i\phi_\mu}$),
\begin{equation}
m_{\tilde{\nu}_1} >  
17\,\text{ TeV}\times (\sin \phi_\mu \tan \beta)^{\frac{1}{2}}.
\end{equation}

\begin{figure}[tbp]
\includegraphics[width=\textwidth]{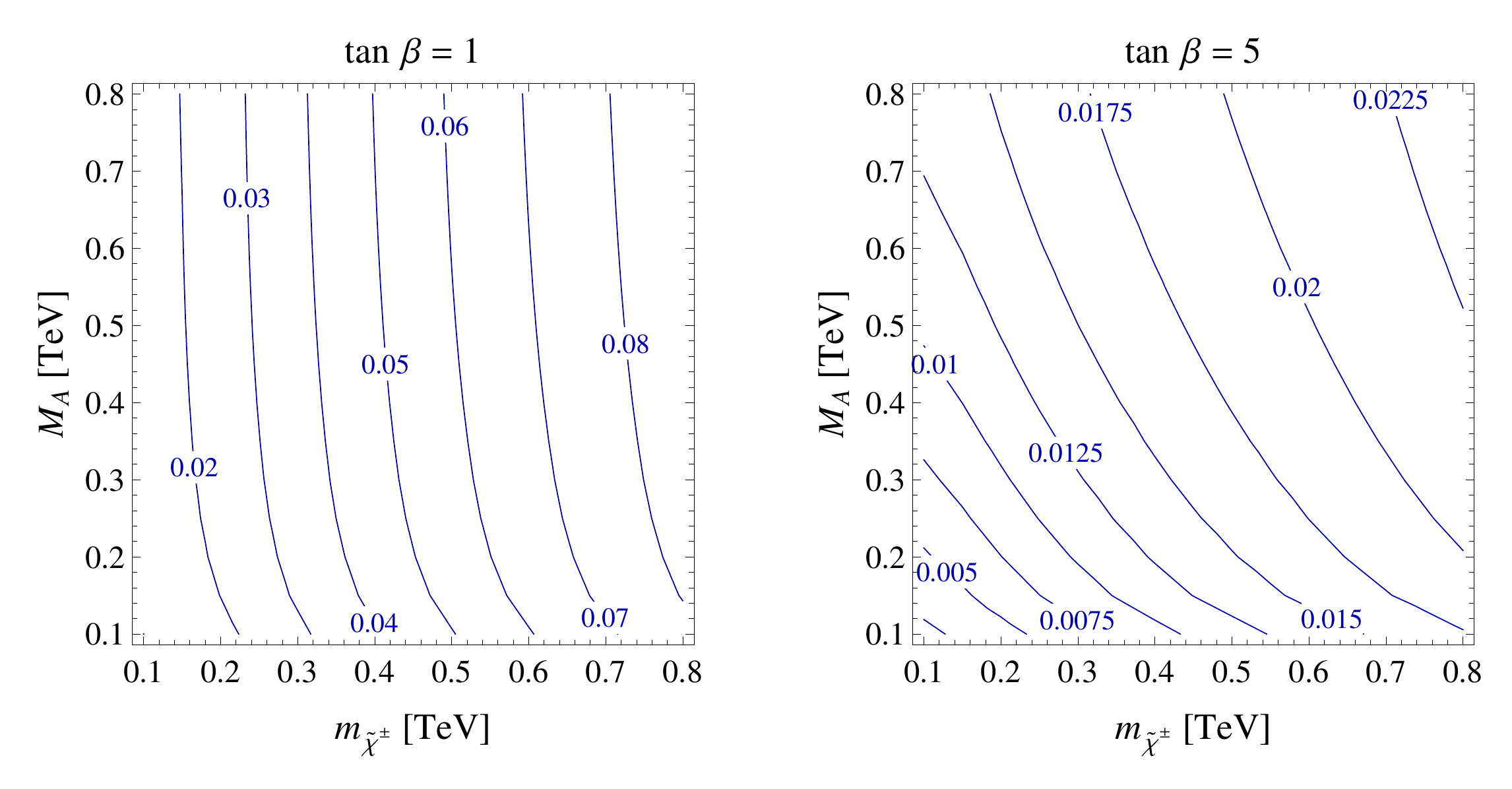}
\caption{Maximum value of the argument of the $\mu$ term, $\phi_\mu$, allowed by two-loop contributions to the electron EDM, in the plane of the common chargino mass $m_{\tilde\chi^\pm}$ and the pseudoscalar Higgs mass $M_A$.}
\label{fig:edm2loop}
\end{figure}

Second, there is a contribution from two-loop Barr-Zee type diagrams involving a chargino loop attached to the electron line by a Higgs and a gauge boson. This contribution is $\tan\beta$ dependent and decouples with the chargino masses. We refer to \cite{Chang:1998uc,Giudice:2005rz,Li:2008kz,Ellis:2008zy,Abel:2001vy} for explicit expressions for these contributions. In fig.~\ref{fig:edm2loop}, we show the constraints on the phase of $\mu$ obtained from these contributions for two different values of $\tan\beta$.
We conclude that, even for very low $\tan\beta$, an $O(1)$ phase of the $\mu$ term is no longer viable with a natural SUSY spectrum that has Higgsinos below a TeV. We stress that this conclusion is independent of the mass of the heavy Higgses, since the light Higgs contribution does not decouple even if they are heavy, as can be seen in the upper part of the left-hand plot in fig.~\ref{fig:edm2loop}.

\section{Conclusions}

The search for the production of new particles, as expected in BSM at the TeV scale, is a  primary task of the next LHC phase. The exploration of most part of the sensitive region of parameter space in motivated theories 
is actually likely to take place in the relatively early stage of the new LHC phase. After that, whatever the findings of this exploration will be, precision measurements in flavour physics (and not only in flavour physics) may play a leading role for a sufficiently  long period of time. With this in mind, we have asked which impact can flavour measurements  have as a whole in the medium term future of particle physics. There is no general way to address this question. Yet the consistency of the CKM description of flavour physics with current data makes interesting to consider an effective picture close enough to the CKM one, based on symmetries and on their breaking patterns only. Under the stated assumptions we see two possible ways to go: $U(3)^3$ (or MFV) and $U(2)^3$ suitably broken.
Needless to say we do not know the origin of these patterns at a fundamental level but we take them as a low energy property of some basic theory at an effective Lagrangian level.

Fig.~\ref{fig:DF2plots1} and eq. (\ref{h_Lambda}) show that current flavour measurements probe scales up to 4 to 7~TeV, depending on the $\Delta F =2$ channel and with the MFV case more restricted. In $\Delta F=1$ channels the constraints are relatively weaker, with the single exceptions of $B\rightarrow X_s \gamma$ and $\epsilon^\prime/\epsilon$. Of the observables shown in fig.~\ref{fig:DF2plots2}, the measurement that should progress more is the one on $\phi^\Delta_s$, but an overall progress is expected on a long series of measurements in the next decade or so  that should reduce the current $20\%$ bound on $h_B$ to about $5\%$ \cite{Charles:2013aka}. In $U(2)^3$ the bound on $h_K$ should go down by about a factor of two, although remaining relatively weaker than the one on $h_B$.

The comparison of the impact of these measurements with the direct searches of new particles can only be made in a specific context. In view of this we have chosen to consider supersymmetry with a flavour structure constrained by symmetries and with a spectrum suggested by naturalness considerations. The main feature of this spectrum, specified in eqs. (\ref{scan_glustop}),\,(\ref{scan_firstmu}), is the heaviness of the squarks of the first two generations. Such a spectrum is at least consistent with a weakly broken $U(2)^3$ symmetry. An interesting feature of supersymmetry and $U(2)^3$ is that some of the virtual exchanges in flavour violating processes are effectively MFV-like whereas only Wino or gluino exchanges can produce a deviation from MFV. 

A scan of the parameter space defined in eqs.~(\ref{scan_glustop})--(\ref{scan_O1}) produces the scatter plots in fig.~\ref{fig:scatter}, controlled by many partial analytic descriptions (see section~\ref{sec:DF2-SUSY-ana}). The ranges of parameters considered extend the regions explorable in direct searches at LHC in its second phase. While all the points are allowed by current direct searches, an important distinction is introduced between points corresponding to a ``compressed" spectrum (light blue) and all other points (dark blue).
Although the vast majority of the points are compatible with the constraints obtained by a general $U(2)^3$ fit of current flavour measurements, the foreseen improvements described above may show a deviation from the SM in flavour physics, most likely corresponding to a spectrum whose lightest fragments should be visible in direct production as well in the second LHC phase. Note in the four upper figures the lines corresponding to MFV.
On the contrary the oblique lines in the two lower figures show the correlations between $s$ and $d$ characteristic of $U(2)^3$, as described in eqs.~(\ref{h-parameters}),\,(\ref{eq:h2obs}). The relatively small deviations from the perfect correlation appearing in these figures is due to the presence of second order effects in the small symmetry breaking parameters that produce a main contribution to the LR operator in $B_s$ mixing (see section~\ref{sec:DF2-SUSY-ana}).

Everywhere in the consideration of possible supersymmetric signals we have taken $\tan{\beta}$ below 5. Also because of this the current constraints on $\Delta F=1$ effects is dominated by the measurement of the $B\rightarrow X_s \gamma$ rate, as shown in fig.~\ref{fig:bsg}. Other measurements that should improve on the current situation are angular CP asymmetries and other observables in $B\rightarrow K^{(*)} \bar{l} l$ decays and, in kaon physics, measurements of $K\rightarrow \pi \nu\bar{\nu} $.

A final point concerns the recently improved limit on the electron EDM. Depending on the value of $\tan{\beta}$ and on the phase $\phi_\mu$ of the $\mu$ parameter, suitably defined, the new bound gives a severe constraint on the sneutrino mass of the first generation. More important for the considerations developed here, however, is the bound coming from two loop diagrams on the masses of the extra Higgs bosons and of charginos, illustrated in fig.~\ref{fig:edm2loop}. A flavour blind phase of order one is not compatible any more with charginos below the TeV range. The separate origin of $\phi_\mu$ from the phases appearing in the flavour violating operators in section~\ref{eff-ops} may explain the absence of signal in the EDM experiments so far. In any way this is quite a significant change with respect to the previous situation.

\section*{Acknowledgements}

D.S.\ thanks Janusz Rosiek for technical support and Thorsten Feldmann for discussions. D.B.\ and F.S.\ thank respectively CERN and the Lawrence Berkeley National Laboratory for hospitality while this work was in the initial stage.
This work was supported by the EU ITN ``Unification in the LHC Era'', 
contract PITN-GA-2009-237920 (UNILHC), by MIUR under contract 2010 YJ2NYW-010, by the DFG cluster of excellence ``Origin and Structure of the Universe'', and by the European Research Council (ERC) under the EU Seventh Framework Programme (FP7/2007-2013) / {\sc Erc} Starting Grant (agreement n. 278234 - ‘{\sc NewDark}’ project) and the ERC Advanced Grant project {\sc Flavour} (agreement n. 267104).

\appendix

\section{Loop functions for meson mixings}

The relevant contributions to the meson mixings of section~\ref{sec:DF2-SUSY-ana}, coming from the box diagrams, can be expressed in terms of the Passarino-Veltman four-point loop functions at vanishing external momentum% (see e.g. \cite{} for explicit expressions)
\begin{align}
D_0(m_1^2, m_2^2, m_3^3,m_4^2) &= \frac{1}{i\pi^2}\int d^4p \frac{1}{(p^2 - m_1^2)(p^2 - m_2^2)(p^2 - m_3^2)(p^2 - m_4^2)},\\
D_2(m_1^2, m_2^2, m_3^3,m_4^2) &= \frac{1}{i\pi^2}\int d^4p \frac{p^2/4}{(p^2 - m_1^2)(p^2 - m_2^2)(p^2 - m_3^2)(p^2 - m_4^2)}.
\end{align}
We define the functions
\begin{equation}
f_{\alpha}(x,y,z) = c_{0,\alpha} m_\alpha^2 D_0(x m_\alpha^2, y m_\alpha^2, z m_\alpha^2, m_\alpha^2) + c_{2,\alpha} D_2(x m_\alpha^2, y m_\alpha^2, z m_\alpha^2, m_\alpha^2),
\end{equation}
where $\alpha = \tilde g, \tilde W, \tilde B, \tilde H^{\pm}, H^{\pm}, W^{\pm}, \tilde g_{(\text{LR})}$, and the coefficients $c_{0,\alpha}, c_{2,\alpha}$ are
\bigskip
\begin{center}
\begin{tabular}{c|ccccccc}
& $\tilde g$ & $\tilde W$ & $\tilde B$ & $\tilde H^\pm$ & $W^\pm$ & $H^\pm$ & $\tilde g_{(\text{LR})}$\\
\hline
$c_0$ & -8/9 & -1/2        & -1/162      &  0       & 2 & 0  & -56/3\\
$c_2$ & -88/9& -5            & -1/81          &  -4       & -2 & -1 & 32/3
\end{tabular}
\end{center}
Writing for convenience $f_{\alpha}(x,y) \equiv f_{\alpha}(x,y,1)$, the contributions to the gluino-induced mixing amplitudes, defined as in \eqref{eq:gluino_K}--\eqref{eq:gluino_Bs}, then read
\begin{align}
F_{1,\tilde g} &= \frac{1}{S_0}\frac{m_W^2}{M_3^2}\frac{\alpha_3^2}{\alpha_2^2}\,K_1\Big(f_{\tilde g}(x_{3},x_{3}) + f_{\tilde g}(x_{1},x_{1}) - 2f_{\tilde g}(x_{1},x_{3})\Big),\\
F_{2,\tilde g} &= \frac{1}{S_0}\frac{m_W^2}{M_3^2}\frac{\alpha_3^2}{\alpha_2^2}\,\left|\frac{\xi_2}{\xi_3}\right| \,K_2\Big(f_{\tilde g}(x_1,x_1) + f_{\tilde g}(x_3,x_2) - f_{\tilde g}(x_2,x_1) - f_{\tilde g}(x_3,x_1)\Big),\\
F_{3,\tilde g} &= \frac{1}{S_0}\frac{m_W^2}{M_3^2}\frac{\alpha_3^2}{\alpha_2^2}\,\left|\frac{\xi_2}{\xi_3}\right|^2 K_3\Big(f_{\tilde g}(x_2,x_2) + f_{\tilde g}(x_1,x_1) - 2f_{\tilde g}(x_2,x_1)\Big),\\
F_{4,\tilde g} &= \frac{1}{S_0}\frac{m_W^2}{M_3^2}\frac{\alpha_3^2}{\alpha_2^2}\, \frac{y_s}{y_b}\, K_4\Big(f_{\tilde g_{(\text{LR})}}(x_3,x_3) + f_{\tilde g_{(\text{LR})}}(x_1,x_1) - 2f_{\tilde g_{(\text{LR})}}(x_3,x_1)\Big),
\end{align}
where $x_i = (m_{Q_i,U_i,D_i}/M_3)^2$, $\xi_a^{ij} = V_{ai}^*V_{aj}$, and $S_0 \simeq 2.31$ is the corresponding SM loop function evaluated at $\mu = m_t$. The factors $K_i$ include the effect of the running and operator mixings \cite{Becirevic:2001jj,Barbieri:2010ar}, as well as the hadronic matrix elements of the relevant operators \cite{Bertone:2012cu,Carrasco:2013zta}, normalised to the SM LL contribution. Analogous expressions hold for the other gaugino contributions.
For the charged Higgs and Higgsino contributions \eqref{eq:FchargedH} one obtains
\begin{align}
F_{H^\pm} &= \frac{1}{S_0}\frac{m_t^4}{m_W^2 m_{H^\pm}^2}K_1\Big(\frac{1}{t_\beta^4} f_{H^\pm}(x_{t},x_{t}) + \frac{1}{t_\beta^2} f_{W^\pm}(x_{t},x_{t},x_{H^\pm})\Big),\\
F_{\tilde H^\pm} & =  \frac{1}{S_0}\frac{m_W^2}{\mu^2} \frac{y_t^4}{g_2^4} \,\frac{1}{s_{\beta}^4}\,K_1 \, f_{\tilde H^\pm}(x_{3},x_{3}).
\end{align}

%#####################################################%
\bibliographystyle{style}
\bibliography{u2susy}
%#####################################################%

\end{document}